\documentstyle[12pt]{article}
\begin{document}
\setlength{\oddsidemargin}{0cm}
\setlength{\baselineskip}{7mm}
\def\beq{\begin{equation}}
\def\eeq{\end{equation}}
\def\beqa{\begin{eqnarray}}
\def\eeqa{\end{eqnarray}}
\def\tr{{\rm tr}}
\def\strut{\rule[-8pt]{0pt}{23pt}}
\newtheorem{lemma}{Lemma}

\begin{titlepage}

    \begin{normalsize}
     \begin{flushright}
                 UT-Komaba/96-30 \\
     		 hep-th/9612148 \\
                 December 1996
     \end{flushright}
    \end{normalsize}
    \begin{LARGE}
       \vspace{1cm}
       \begin{center}
         D-Branes on Group Manifolds \\
       \end{center}
    \end{LARGE}

  \vspace{5mm}

\begin{center}
           Mitsuhiro K{\sc ato}
           \footnote{E-mail address:
              kato@hep1.c.u-tokyo.ac.jp}
                \ \  and \ \
           Tomoharu O{\sc kada}
           \footnote{E-mail address:
              okada@hep1.c.u-tokyo.ac.jp} \\
      \vspace{4mm}
        {\it Institute of Physics} \\
        {\it University of Tokyo, Komaba}\\
        {\it Meguro-ku, Tokyo 153, Japan}\\
      \vspace{1cm}

    \begin{large} ABSTRACT \end{large}
        \par
\end{center}
\begin{quote}
 \begin{normalsize}
Possible Dirichlet boundary states for WZW models with untwisted affine super
Kac-Moody symmetry are classified for all compact simple Lie groups. They are
obtained by inner- and outer-automorphism of the group. D-brane world-volume
turns out to be a group manifold of a symmetric subgroup, so that the moduli
space of D-brane is a irreducible Riemannian symmetric space. It is also
clarified how these D-branes are transformed to each other under abelian
T-duality of WZW model. Our result implies, for example, there is no D-particle
on the compact simple group manifold. When the D-brane world-volume contains
$S^1$ factor, the D-brane moduli space becomes hermitian symmetric space and
the open string world-sheet instantons are allowed.
 \end{normalsize}
\end{quote}

\end{titlepage}
\vfil\eject

\section{Introduction}

It is now widely recognized that the D-branes are key ingredients of the
non-perturbative physics of the string theories \cite{P}. A conformal field
theoretic (CFT) treatment of the interaction of the closed string with the
background D-brane is described by the boundary state adapted to Dirichlet
boundary condition. However, most of the CFT analysis so far are restricted to
the flat D-brane, except a few attempts \cite{OOY}. In order to get more
insights on the D-brane dynamics and eventually to achieve a full quantum
treatment of the D-branes themselves\footnote{For D-particle, an effort toward
this end is in ref.\cite{HK}}, it is valuable for us to experience various
situations such as a curved D-brane in a curved space.

In the present paper, we will study Dirichlet boundary states which describe
D-branes living in the group manifolds \cite{CS}. As is well known, strings on
the group manifolds are described by the WZW models, which possess left and
right affine (super or non-super) Kac-Moody algebras \cite{GW}. The condition
imposed to the boundary state is given in terms of the Kac-Moody currents. For
the purely Neumann boundary condition, the building blocks of the associated
boundary state has been known as the Ishibashi \cite{I} state. We will
generalize it to accommodate the Dirichlet boundary condition. For the abelian
currents, one can take an arbitrary number of Dirichlet directions. It is,
however, not the case for the non-abelian currents; non-trivial commutation
relations of the currents require consistency among the boundary conditions for
diverse directions. We will classify possible Dirichlet boundary conditions
with respect to super and non-super untwisted Kac-Moody currents for all
compact simple Lie groups.

T-duality of the WZW model transforms one boundary state to another. We will
show how the boundary states thus obtained are transformed to each other under
the abelian T-duality.

It is known that under abelian T-duality non-trivial monodromy for world-sheet
genus greater than one can be induced for the dual theory, when the level of
the Kac-Moody algebra is greater than one \cite{G}. That means twisted sectors
may appear in the dual theory even if the original theory consists only of
untwisted sector. Thus we need to be equipped with the boundary states for the
twisted Kac-Moody currents. In the present paper, however, we restrict our
study only to the untwisted sector. Our argument can be applied
straightforwardly to the case where left and right states have the same twist,
whereas the asymmetric twist case should be considered separately. We will
leave the latter case for a future publication.

The organization of the paper is the following. We give a criterion for the
consistent Dirichlet boundary condition for the WZW models in the subsequent
section. A generalized version of the Ishibashi state is also constructed
there. In section 3, we make a classification of the Dirichlet boundary state
for all compact simple Lie groups. Abelian T-duality transformation of these
boundary states are discussed in section 4. The last section is devoted to
discussions and comments. Some technical points are explained in two
appendices.

\section{Boundary states in WZW model}

Let us begin by recalling the boundary states in flat space or abelian current
case \cite{CLNY,PC}. Denoting $\alpha_n$ and $\tilde{\alpha}_n$ for the closed
string left and right oscillators in a standard way, Neumann and Dirichlet
boundary states are defined by the following conditions:
\beqa
{\rm Neumann:}&\qquad &(\alpha_n + \tilde{\alpha}_{-n})|N\rangle = 0\,, \\
{\rm Dirichlet:}&\qquad &(\alpha_n - \tilde{\alpha}_{-n})|D\rangle = 0\,.
\eeqa
The solutions for these equations are well known as
\beqa
|N\rangle &=& \int dp\, f(p)\, \exp(-\sum_{n>0}
\frac{1}{n}\alpha_{-n}\tilde{\alpha}_{-n})
	|p\rangle\otimes\widetilde{|-p\rangle}\,, \\
|D\rangle &=& \int dp\, f(p)\, \exp(+\sum_{n>0}
\frac{1}{n}\alpha_{-n}\tilde{\alpha}_{-n})
	|p\rangle\otimes\widetilde{|p\rangle}\,,
\eeqa
where $|p\rangle$ and $\widetilde{|p\rangle}$ are the eigenstates of $\alpha_0$
and $\tilde{\alpha}_0$, respectively, with eigenvalue $p$ and $f(p)$ is an
arbitrary function.

Our aim in this section is to extend these boundary states to the case of the
non-abelian affine super and non-super Kac-Moody algebras. We consider the
algebras associated to the compact simple Lie group $G$, for brevity. But the
basic part of our argument can be easily extended to more general groups.

Denoting left and right (non-super for the moment) Kac-Moody currents by $J^a$
and $\tilde{J}^a$ ($a=1,2,\cdots,d=\dim(G)$) respectively, a generalization of
the above boundary condition is
\beqa
{\rm Neumann:}&\qquad &(J^a_n + \tilde{J}^a_{-n})|B\rangle = 0\,, \\
{\rm Dirichlet:}&\qquad &(J^a_n - \tilde{J}^a_{-n})|B\rangle = 0\,.
\eeqa
It is, however, not possible to impose arbitrarily Neumann or Dirichlet
condition. Since the currents satisfy non-trivial algebra
\beqa
[ J^a_n \,, J^b_m ] &=& if^{abc} J^c_{n+m} +
	\frac{k}{2}n\delta^{ab}\delta_{n+m,0}\,, \\
{[ \tilde{J}^a_n \,, \tilde{J}^b_m ]} &=& if^{abc} \tilde{J}^c_{n+m} +
	\frac{k}{2}n\delta^{ab}\delta_{n+m,0}\,, \\
{[ J^a_n \,, \tilde{J}^b_m ]} &=& 0\,,
\eeqa
the boundary condition has to be consistent with the algebra. For example, if
we impose Dirichlet condition in two directions, say $a$ and $b$, then the
relation
\beq
[ J^a_n - \tilde{J}^a_{-n} \,, J^b_m - \tilde{J}^b_{-m} ] |B\rangle
	= i f^{abc}( J^c_{n+m} + \tilde{J}^c_{-n-m} ) |B\rangle
\eeq
forces us to impose Neumann conditions to the direction $c$ of non-vanishing
$f^{abc}$. It is clear, in particular, that we cannot put Dirichlet conditions
in all the directions. That means there is no D-particle on the compact simple
group manifold.

We shall look for consistent sets of boundary conditions in the following form:
\beq
\left( J^a_n + \tau(\tilde{J}^a_{-n}) \right) |B\rangle = 0\,, \label{BC}
\eeq
where $\tau$ is a certain map of the current. It is easily seen that the
consistency of this condition requires a relation
\beq
[ \tau(\tilde{J}^a_n)\,, \tau(\tilde{J}^b_m) ] =
	\tau([\tilde{J}^a_n\,,\tilde{J}^b_m])\,,
\eeq
so that the $\tau$ should be an automorphism of the algebra. Since we are
concerned with Dirichlet and Neumann condition, we only consider automorphism
of the type
\beq
\tau(\tilde{J}^a_n) = U^{ab} \tilde{J}^b_n\,, \label{U}
\eeq
with a orthogonal matrix $U$.\footnote{Lattice translation in the affine Weyl
group, which mixes the generators on the different level, cannot be written in
this form, but does not give a simple Dirichlet or Neumann condition.} The
orthogonality of $U$ guarantees the invariance of the $\tr(T^aT^b)$, where
$T^a$'s are the generators of the corresponding non-affine Lie algebra. The
condition that eq.(\ref{U}) gives automorphism is
\beq
U^{ad}U^{be}f^{def} = f^{abc}U^{cf}\,. \label{UUF}
\eeq
This is equivalent to the invariance of $\tr([T^a,T^b]T^c)$. One can easily see
$\tau(\tilde{L}_n)=\tilde{L}_n$ for the Virasoro generator given in the
Sugawara form in terms of $\tilde{J}^a_n$. Hence our boundary condition is
automatically consistent with the conformal invariance
\beq
(L_n-\tilde{L}_{-n})|B\rangle = 0\,.
\eeq

None of the above argument is changed in the case of super Kac-Moody algebra:
\beqa
[ J^a_n\,,J^b_m ]&=&if^{abc}J^c_{n+m} +
	\frac{k}{2}n\delta^{ab}\delta_{n+m,0}\,, \\
{[ J^a_n\,,j^b_m ]}&=&if^{abc}j^c_{n+m}\,, \\
\{ j^a_n\,,j^b_m \}&=&\frac{k}{2}\delta^{ab}\delta_{n+m,0}\,,
\eeqa
where $j^a_n$ is a fermionic partner of the current $J^a_n$, and the same
algebra is satisfied by the right moving current $\tilde{J}^a_n$ and its
fermionic partner $\tilde{j}^a_n$. We take the boundary condition as
\beqa
\left( J^a_n + \tau(\tilde{J}^a_{-n}) \right) |B\rangle = 0\,, \label{BCJ}\\
\left( j^a_n \pm i\tau(\tilde{j}^a_{-n}) \right) |B\rangle = 0\,.
\eeqa
If we take identity map for $\tau$, then this reduces to the usual Neumann
boundary condition. As in the non-super case, $\tau$ should be an automorphism
of the super algebra, and we consider the one of the type given by
\beqa
\tau(\tilde{J}^a_n) &=& U^{ab} \tilde{J}^b_n\,, \\
\tau(\tilde{j}^a_n) &=& U^{ab} \tilde{j}^b_n\,.
\eeqa
Here the orthogonal matrix $U$ is common to $\tilde{J}$ and $\tilde{j}$ because
of supersymmetry, and the condition of automorphism is given by the same
eq.(\ref{UUF}) as before.
Since $\tau(\tilde{L}_n)=\tilde{L}_n$ and $\tau(\tilde{G}_n)=\tilde{G}_n$ for
super-Virasoro generators $\tilde{L}_n$ and $\tilde{G}_n$, they are consistent
with the superconformal invariance
\beqa
(L_n-\tilde{L}_{-n})|B\rangle &=& 0\,, \\
(G_n\mp i\tilde{G}_{-n})|B\rangle &=& 0\,.
\eeqa
Thus our problem for both super and non-super algebra is to find possible
orthogonal matrices $U$ satisfying eq.(\ref{UUF}).

Before going into the classification of automorphisms, let us touch on the
construction of the boundary state. For purely Neumann case, the building block
of the boundary state was constructed by Ishibashi \cite{I}:
\beq
|w\rangle\rangle = \sum_n |w,n\rangle \otimes \Theta \widetilde{|w,n\rangle}\,,
\eeq
where $|w,n\rangle$ and $\widetilde{|w,n\rangle}$ are complete orthogonal basis
of a representation $w$ for left and right current algebra, and $\Theta$ is an
anti-unitary operator which satisfies $\Theta\tilde{J}^a_n\Theta^{-1} =
-\tilde{J}^a_n$ and $\Theta\tilde{j}^a_n\Theta^{-1} = \mp i\,\tilde{j}^a_n$.
For our general boundary condition, it should be slightly modified:
\beq
|w\rangle\rangle = \sum_n |w,n\rangle \otimes
	{\cal T} \Theta \widetilde{|w,n\rangle}\,, \label{BS}
\eeq
where ${\cal T}$ is an unitary operator which generates an automorphism ${\cal
T}\tilde{J}^a_n{\cal T}^{-1} = \tau(\tilde{J}^a_n)$ and ${\cal
T}\tilde{j}^a_n{\cal T}^{-1} = \tau(\tilde{j}^a_n)$. We can prove that the
state (\ref{BS}) satisfies the condition (\ref{BCJ}) as almost same way as
Ishibashi state does Neumann condition. Let us take an arbitrary bra-state
$\langle i|\otimes\widetilde{\langle j|}$ in the representation space of $w$.
Then we can show
\beqa
\langle i|\otimes\widetilde{\langle
j|}(J^a_n\!\!&+&\!\!\tau(\tilde{J}^a_{-n}))|w\rangle\rangle\nonumber\\
&=&\sum_n\langle i|J^a_n|w,n\rangle
\widetilde{\langle j|} {\cal T} \Theta \widetilde{|w,n\rangle}
+\sum_n\langle i|w,n\rangle
\widetilde{\langle j|}\tau(\tilde{J}^a_{-n}){\cal T} \Theta
\widetilde{|w,n\rangle} \nonumber\\
&=&\sum_n\langle i|J^a_n|w,n\rangle
\widetilde{\langle j|} {\cal T} \Theta \widetilde{|w,n\rangle}
-\sum_n\langle i|w,n\rangle
\widetilde{\langle j|}{\cal T} \Theta \tilde{J}^a_{-n}\widetilde{|w,n\rangle}
\nonumber\\
&=&\sum_n\langle i|J^a_n|w,n\rangle
\widetilde{\langle w,n|}\Theta^{\dag} {\cal T}^{\dag}\widetilde{|j\rangle}
-\sum_n\langle i|w,n\rangle
\widetilde{\langle w,n|}\tilde{J}^a_{n}\Theta^{\dag} {\cal T}^{\dag}
\widetilde{|j\rangle} \nonumber\\
&=&\sum_n\widetilde{\langle i|}\tilde{J}^a_n\widetilde{|w,n\rangle}
\widetilde{\langle w,n|}\Theta^{\dag} {\cal T}^{\dag}\widetilde{|j\rangle}
-\sum_n\widetilde{\langle i|}\widetilde{w,n\rangle}
\widetilde{\langle w,n|}\tilde{J}^a_{n}\Theta^{\dag} {\cal T}^{\dag}
\widetilde{|j\rangle} \nonumber\\
&=&0\,.
\eeqa
Since the state $\langle i|\otimes\widetilde{\langle j|}$ is arbitrary, $\left(
J^a_n+\tau(\tilde{J}^a_{-n})\right)|w\rangle\rangle = 0$ holds. Similarly, we
can show $\left( j^a_n\pm i\tau(\tilde{j}^a_{-n})\right)|w\rangle\rangle = 0$.
Thus a general boundary state is given by their linear combination $|B\rangle =
\sum_w c(w) |w\rangle\rangle$.

\section{Classification of Dirichlet boundary states}

Now we are to classify possible automorphism which gives consistent
boundary condition. From the results in the previous section, it is sufficient
to consider the automorphism $\tau(T^a)=U^{ab}T^b$ of non-affine version of
the algebra
\beq
[T^a,T^b]=if^{abc}T^c\,.
\eeq
Since we are interested in the Dirichlet or Neumann
condition, not its mixture in one direction, our orthogonal matrix $U^{ab}$
should satisfy $U^2={\bf 1}$.\footnote{If we use an automorphism other than
${\bf Z}_2$, corresponding boundary condition we get is a mixed
Dirichlet-Neumann condition which can be interpreted as the boundary coupled
with open string background fields.} This yields $U$ to be a symmetric matrix.
More explicitly, if we take an appropriate hermitian
basis, $U$ becomes a diagonal matrix whose entries are $\pm 1$. Then in
this basis we can divide all the generators into two sets; $S=\{T^a |\,
\tau(T^a)=T^a\}$ and $A=\{T^a |\, \tau(T^a)=-T^a\}$. They satisfy
\beq
[S,S]\subset S\,,\qquad [S,A]\subset A\,,\qquad [A,A]\subset S\,.
\eeq
Thus $S$ forms subalgebra which is known as {\em symmetric
subalgebra\/}. So our task is now reduced to the classification of the
symmetric subalgebras. Though the answer is already known \cite{S}, we
want not just a list of the subalgebra but their mutual relations under
T-duality. Therefore, we go into a fine structure of the automorphism which
will be relevant for the argument in
section 4.

There are two classes of the automorphism for the compact simple Lie
algebra: inner-automorphism and outer-automorphism. In order to study
these, it is convenient to use Cartan-Weyl basis instead of hermitian
basis. Let us denote Cartan generators by $H^i$
($i=1,2,\cdots,r={\rm rank}(G)$) and a generator associated to a root $\alpha$
by $E^{\alpha}$.

Firstly we study inner-automorphism, which is given by an adjoint action
\beq
\tau(T^a)=g\,T^ag^{-1}\,,\qquad g\in G\,.
\eeq
It is general enough to take $g$ as being made only of the Cartan
generators \cite{GO}: $g=\exp(i\eta\cdot H)$. Then each generator is
transformed under this automorphism as
\beqa
\tau(H^i)&=&H^i\,,\\
\tau(E^{\alpha})&=&\exp(i\eta\cdot\alpha)\, E^{\alpha}\,.\label{EA}
\eeqa
Since every positive (resp.~negative) root is expressed by a linear sum of
simple roots
$\alpha_i$ ($i=1,2,\cdots,{\rm rank}(G)$) with non-negative (non-positive)
integer
coefficients,
the phase appeared in eq.(\ref{EA}) for a positive (negative) root is written
as a product of the phases for
simple roots with non-negative (non-positive) integer power. So the independent
freedom of our choice is in a set of
$\{\epsilon_i=\exp(i\eta\cdot\alpha_i)\}$. Possible values of $\epsilon_i$ are
necessarily $\pm 1$ from the property $U^2={\bf 1}$. For a generic root
$\alpha=\sum_im_i\alpha_i$, the phase factor becomes a product
$\prod_i\epsilon_i{}^{m_i}$.

It is useful to define a quantity
\beq
T_G(\{\epsilon_i\})=\sum_{\alpha\in\Delta_+}\exp(i\eta\cdot\alpha)
	=\sum_{\sum_i m_i\alpha_i\,\in\Delta_+}
	\prod_{i=1}^{{\rm rank}(G)}\epsilon_i{}^{m_i}\,,
\eeq
where $\Delta_+$ is a set of all positive roots. If we denote $N_{\pm}$ as the
number of positive root generators $E^{\alpha}$ whose phase factor
$\exp(i\eta\cdot\alpha)$ is $\pm1$ respectively, then $T_G$ gives $N_+-N_-$
under a given $\{\epsilon_i\}$. Thereby the number of Dirichlet directions
$N_D$, {\it i.e.}\ $2N_-$, is given by
\beq
N_D = \frac{1}{2}(d-r)-T_G\,,\label{ND}
\eeq
where $d=\dim(G)$ and $r={\rm rank}(G)$.

On the other hand, the number of Neumann direction is the dimension of the
D-brane world-volume embedded in the group manifold $G$. It is clear from the
above argument that the D-brane world-volume itself is a group manifold
associated to the symmetric subgroup which we denote $H$.

We examined all possible choices of $\{\epsilon_i\}$ and determined $N_D$ and
$H$ for all compact simple Lie groups. The result is shown in
Table~\ref{Inner}. We omitted in the table the identity automorphism which
gives $N_D=0$ and $H=G$, {\it i.e.}\ pure Neumann boundary condition. Some
technical statements are in the Appendix~A.

\begin{table}
\begin{center}
\begin{tabular}{||l|l|l||} \hline
\hfil$G$ & \hfil$H$ & \hfil$N_D$ \strut\\\hline
$A_n = SU(n+1)$ & $A_{n-k}\otimes A_{k-1}\otimes U(1)$ &
    $2k(n+1-k),\quad k=1,2,\cdots,[\frac{n+1}{2}]$ \strut\\
    \hfill$d=n(n+2)$ & & \strut\\\hline
$B_n = SO(2n+1)$ & $B_{n-k}\otimes D_k$ &
    $2k(2n+1-2k),\quad k=1,2,\cdots,n-1$ \strut\\\cline{2-3}
    \hfill$d=n(2n+1)$ & $D_n$ & $2n$ \strut\\\hline
$C_n = Sp(n)$ & $A_{n-1}\otimes U(1)$ & $n(n+1)$ \strut\\\cline{2-3}
    \hfill$d=n(2n+1)$ & $C_{n-k}\otimes C_k$ &
    $4k(n-k),\quad k=1,2,\cdots,[\frac{n}{2}]$ \strut\\\hline
$D_n = SO(2n)$ & $A_{n-1}\otimes U(1)$ & $n(n-1)$ \strut\\\cline{2-3}
    \hfill$d=n(2n-1)$ & $D_{n-k}\otimes D_k$ &
    $4k(n-k),\quad k=1,2,\cdots,[\frac{n}{2}]$ \strut\\\hline
$E_6$ \hfil$d=78$ & $D_5\otimes U(1)$ & $32$ \strut\\\cline{2-3}
    &\strut $A_5\otimes A_1$ & $40$ \\\hline
$E_7$ \hfil$d=133$ & $E_6\otimes U(1)$ & $54$ \strut\\\cline{2-3}
    & $D_6\otimes A_1$ & $64$ \strut\\\cline{2-3}
    & $A_7$ & $70$ \strut\\\hline
$E_8$ \hfil$d=248$ & $E_7\otimes A_1$ & $112$ \strut\\\cline{2-3}
    & $D_8$ & $128$ \strut\\\hline
$F_4$ \hfil$d=52$ & $B_4$ & $16$ \strut\\\cline{2-3}
    & $C_3\otimes A_1$ & $28$ \strut\\\hline
$G_2$ \hfil$d=14$ & $A_1\otimes A_1$ & $8$ \strut\\\hline
\end{tabular}
\end{center}
\caption{List of symmetric subgroup $H$ other than $G$ itself obtained
    by inner-automorphism. $N_D$ is the number of Dirichlet directions.}
\protect\label{Inner}
\end{table}

We note here that, for the inner-automorphism case, the symmetric subgroup is a
maximal regular subgroup. Smaller (non-maximal) subgroup cannot be obtained by
${\bf Z}_2$ ({\it i.e.}\ $U^2={\bf 1}$) automorphism. This is followed by that
if we consider multiple D-branes they are likely to intersect.

Next we turn to the study of outer-automorphism. Outer-automorphism is achieved
by the automorphism of Dynkin diagram. Therefore only $A_n$, $D_n$ and $E_6$
are relevant. Under the automorphism each generator is transformed as
\beqa
\tau(\alpha\cdot H)&=&\tau(\alpha)\cdot H\,,\\
\tau(E^{\alpha})&=&\varepsilon_{\alpha}E^{\tau(\alpha)}\,,
\eeqa
where $\varepsilon_{\alpha}=\varepsilon_{-\alpha}$ is either $+1$ or $-1$
consistently with the algebra. Note that if $\alpha\in\Delta_+$ then
$\tau(\alpha)\in\Delta_+$.
The phase factor $\varepsilon_{\alpha}$ satisfies the following properties
(proof is given in Appendix~B):
\begin{lemma}\
\begin{enumerate}
\item $\varepsilon_{\alpha}\varepsilon_{\tau(\alpha)}=1$.
\item If $\alpha\neq\tau(\alpha)$ and
      $\alpha , \alpha+\tau(\alpha) \in\Delta_+$,
      then $\varepsilon_{\alpha+\tau(\alpha)}=-1$.
\item If $\alpha\neq\tau(\alpha)$, $\beta=\tau(\beta)$ and
      $\alpha ,\beta ,\alpha+\beta ,\tau(\alpha)+\beta ,
      \alpha+\beta+\tau(\alpha) \in \Delta_+$
      but $\alpha+\tau(\alpha)\not\in\Delta_+$,
      then $\varepsilon_{\alpha+\beta+\tau(\alpha)}=\varepsilon_{\beta}$.
\end{enumerate}
\end{lemma}

Let us denote a simple root with property $\tau(\alpha_i)=\alpha_i$ by
$\alpha_{i_{\|}}$ and one with $\tau(\alpha_i)\neq\alpha_i$ by
$\alpha_{i_{\perp}}$. Then apparently
$(\alpha_{i_{\perp}}-\tau(\alpha_{i_{\perp}}))\cdot H$ belongs to the set $A$,
and the rest of Cartan generators do to symmetric subalgebra $S$. As for the
root generator $E^{\alpha}$ with $\alpha\neq\tau(\alpha)$,
\beqa
E^{\alpha}+\varepsilon_{\alpha}E^{\tau(\alpha)}&\in S\,,\\
E^{\alpha}-\varepsilon_{\alpha}E^{\tau(\alpha)}&\in A\,,
\eeqa
while $E^{\alpha}$ with $\alpha=\tau(\alpha)$ depends on its phase
$\varepsilon_{\alpha}$. For the latter, invariance of the algebra restricts
arbitrariness. Since $\alpha\in\Delta_+$ can be written as
\beq
\alpha=\sum m_{i_{\|}}\alpha_{i_{\|}} +
  \sum m_{i_{\perp}}\left(\alpha_{i_{\perp}}+\tau(\alpha_{i_{\perp}})\right)
  \qquad {\rm for }\qquad \alpha=\tau(\alpha)\,,
\eeq
the above lemma lead us to the relation:
\beq
\varepsilon_{\alpha}=\left\{
  \begin{array}{ll}
    -1 & \mbox{if all $m_{i_{\|}}=0$} \\
    \prod_{i_{\|}}\varepsilon_{\alpha_{i_{\|}}}{}^{m_{i_{\|}}}
    & \mbox{otherwise}
  \end{array}
  \right.\qquad {\rm for}\qquad \alpha=\tau(\alpha)\,.
\eeq

This looks like the relation of the phases encountered in the
inner-automorphism case. Actually, if we set $\varepsilon_{\alpha_{i_{\|}}}=1$
for all invariant simple roots as a representative, then each of the other
choices is expressed as a product of this representative automorphism and an
inner-automorphism with $\epsilon_{i_{\|}}=\varepsilon_{\alpha_{i_{\|}}}$ and
$\epsilon_{i_{\perp}}=1$ which commutes with the representative. We show in
Table~\ref{Outer} the list of symmetric subgroup thus obtained. For each $G$,
the first $H$ is the representative. Note that the symmetric subgroup obtained
by outer-automorphism is a maximal special subgroup \cite{S}.

\begin{table}
\begin{center}
\begin{tabular}{||l|l|l||} \hline
\hfil$G$ & \hfil$H$ & \hfil$N_D$ \strut\\\hline
$A_{2l}$ & $B_l$ & $2l^2+3l$ \strut\\\hline
$A_{2l-1}$ & $C_l$ & $2l^2-l-1$ \strut\\\cline{2-3}
    & $D_l$ & $2l^2+l-1$ \strut\\\hline
$D_n$ & $B_{n-1}$ & $2n-1$ \strut\\\cline{2-3}
    & $B_{n-k-1}\otimes B_k$ &
    $2n-1+4k(n-k-1),\quad k=1,2,\cdots,[\frac{n-1}{2}]$ \strut\\\hline
$E_6$ & $F_4$ & $26$ \strut\\\cline{2-3}
    &\strut $C_4$ & $42$ \\\hline
\end{tabular}
\end{center}
\caption{List of symmetric subgroup $H$ obtained by outer-automorphism.
    $N_D$ is the number of Dirichlet directions.}
\protect\label{Outer}
\end{table}

Combining the Table~\ref{Inner} and~\ref{Outer} we have obtained all the
symmetric subgroup $H$ of compact simple Lie group $G$. To make a distinction
between inner- and outer-automorphism is important for the study of abelian
T-duality transformation of these boundary states which is the subject of the
next section.

\section{Abelian T-duality transformations}

An abelian T-duality transformation in the WZW model is nothing but an
automorphism on the left and/or right Kac-Moody currents \cite{K,AAL}:
\beqa
J^a_n &\rightarrow & L^{ab}J^b_n\,,\\
\tilde{J}^a_n &\rightarrow & R^{ab}\tilde{J}^b_n\,,
\eeqa
where the matrices $L$ and $R$ are orthogonal matrices which satisfy the same
relation of eq.(\ref{UUF}) as $U$. For the super Kac-Moody case, they are
supplemented by the transformation on the partners with the same matrices
\beqa
j^a_n &\rightarrow & L^{ab}j^b_n\,,\\
\tilde{j}^a_n &\rightarrow & R^{ab}\tilde{j}^b_n\,.
\eeqa

Under these transformation, our general boundary condition is transformed to
\beqa
\left( L^{ab}J^b_n + U^{ab}R^{bc}\tilde{J}^c_{-n} \right) |B\rangle = 0\,,\\
\left( L^{ab}j^b_n \pm iU^{ab}R^{bc}\tilde{j}^c_{-n} \right) |B\rangle = 0\,.
\eeqa
In use of an appropriate basis change, this is rewritten as
\beqa
\left( J^a_n + (URL^{T})^{ab}\tilde{J}^b_{-n} \right) |B\rangle = 0\,,\\
\left( j^a_n \pm i(URL^{T})^{ab}\tilde{j}^b_{-n} \right) |B\rangle = 0\,.
\eeqa
Hence the boundary condition of a matrix $U$ is transformed to that of
$URL^{T}$. Especially, pure Neumann condition is transformed to the boundary
condition of $RL^{T}$. If we require this to be one of our Dirichlet
conditions, then the matrix $V=RL^{T}$ should satisfy $V^2={\bf 1}$, which we
assume in the following. The matrix $V$ satisfies all conditions that $U$
satisfies: orthogonality, eq.(\ref{UUF}) and $V^2={\bf 1}$. Therefore $V$ falls
into our classification of $U$.

Now we investigate how the boundary conditions of $U$ are related to each other
by the transformation $V$.
Since $(UV)^2={\bf 1}$ is satisfied if and only if $[U,V]=0$, $U$ should be
diagonal in the basis in which $V$ is diagonal. Let us fix the basis in this
way. Obviously, there are two classes, one with and without outer-automorphism.
We shall call them inner class and outer class respectively.

Firstly we consider the inner class, which we denote $I_G$. This is a whole set
of possible $U$ via inner-automorphism for a given $G$, {\it i.e.} the listed
in Table~\ref{Inner} plus identity, all diagonalized in the same basis. Each
element in $I_G$ is uniquely labeled by $\{\epsilon_i\}$ and a product of two
elements are defined by
\beq
U_{\{\epsilon_i^{(1)}\}} U_{\{\epsilon_i^{(2)}\}} =
	U_{\{\epsilon_i^{(1)}\epsilon_i^{(2)}\}}\,.\label{prodI}
\eeq
The corresponding symmetric subgroup is identified by $N_D$ using the formula
(\ref{ND}). Hence, if a transformation $V$ belongs to $I_G$, then a boundary
condition $U\in I_G$ is transformed to $UV\in I_G$ according to the product
rule (\ref{prodI}).

Next we turn to the outer class, which we denote $O_G$. This is a whole set of
possible $U$ via outer-automorphism for a given $G$ (Table~\ref{Outer}) and all
$U\in I_G$ which commute with the matrices of outer-automorphism. The latter
forms a closed subset (we denote it $\hat{I}_G$) whose element is uniquely
labeled by $\{\varepsilon_{\alpha_{i_{\|}}}\}$.  The product rule in
$\hat{I}_G$ is similar to the rule (\ref{prodI}):
\beq
U_{\{\varepsilon_{\alpha_{i_{\|}}}^{(1)}\}}
	U_{\{\varepsilon_{\alpha_{i_{\|}}}^{(2)}\}} =
	U_{\{\varepsilon_{\alpha_{i_{\|}}}^{(1)}
	\varepsilon_{\alpha_{i_{\|}}}^{(2)}\}}\,.\label{prodO1}
\eeq
If a matrix $\hat{U}$ corresponds to the representative of outer-automorphism,
then $O_G$ consists of $U_{\{\varepsilon_{\alpha_{i_{\|}}}\}}$ and
$\hat{U}_{\{\varepsilon_{\alpha_{i_{\|}}}\}} =
\hat{U}U_{\{\varepsilon_{\alpha_{i_{\|}}}\}}$. The product rule in $O_G$ is
given by the eq.(\ref{prodO1}) combined with
\beqa
\hat{U}_{\{\varepsilon_{\alpha_{i_{\|}}}^{(1)}\}}
	U_{\{\varepsilon_{\alpha_{i_{\|}}}^{(2)}\}} =
	\hat{U}_{\{\varepsilon_{\alpha_{i_{\|}}}^{(1)}
	\varepsilon_{\alpha_{i_{\|}}}^{(2)}\}}\,,\label{prodO2}\\
\hat{U}_{\{\varepsilon_{\alpha_{i_{\|}}}^{(1)}\}}
	\hat{U}_{\{\varepsilon_{\alpha_{i_{\|}}}^{(2)}\}} =
	U_{\{\varepsilon_{\alpha_{i_{\|}}}^{(1)}
	\varepsilon_{\alpha_{i_{\|}}}^{(2)}\}}\,.\label{prodO3}
\eeqa
As is for the inner class, a boundary condition $U\in O_G$ is transformed to
$UV\in O_G$ by a transformation $V\in O_G$ according to the product rule
(\ref{prodO1})$\sim$(\ref{prodO3}).

Thus we have shown mutual relationship of the general boundary conditions under
the abelian T-duality of the WZW model which satisfies $(RL^{T})^2={\bf 1}$. If
we loosen this condition, we have to include boundary conditions with $U$ such
that $U^2\neq {\bf 1}$. That means we need to argue the boundary coupled to the
 open string background, which is out of scope of the present paper.

\section{Discussions}

We have shown that the general Dirichlet boundary states of the WZW model can
be constructed by the ${\bf Z}_2$ automorphism of the group so that the D-brane
world-volume is the group manifold of the symmetric subgroup. The coordinates
in the directions perpendicular to the D-brane world-volume corresponds to the
moduli of the D-brane. Therefore the moduli space of the D-brane on group
manifold is a symmetric space $G/H$ given by the target group manifold $G$
divided by the world-volume $H$. Indeed Table~\ref{Inner} and~\ref{Outer}
correspond essentially to the Cartan's classification of the irreducible
Riemannian symmetric spaces \cite{H}.

It is known that, when the symmetric subgroup $H$ contains $U(1)$ or $SO(2)$
factor, the resulting $G/H$ becomes hermitian symmetric space. We do have such
cases as is seen from Table~\ref{Inner}. In this case, we can construct the
$N=2$ superconformal generators from the defining super Kac-Moody currents
\cite{KS}. Since they are related to the geometrical information of the D-brane
moduli space, some interesting implications for the D-brane dynamics may be
expected, though we have not fully clarified yet.

Related to this, we should also note that the open string world-sheet
instantons are allowed for the above case. Let us consider a disk, for
instance. Its boundary, which is $S^1$, should be attached to the D-brane
world-volume $H$. Then the non-trivial homotopy $\pi_1(H)=\pi_1(U(1))={\bf Z}$
implies the existence of the world-sheet instantons, which is smooth everywhere
on the disk if $\pi_1(G)=0$. For simple $G$, instantons and anti-instantons
contribute to the disk amplitude with equal weight.

We have not dealt with twisted sectors in the present paper. As we mentioned in
the introduction, our argument can be directly applicable to the case where the
left and right sectors have same twist. Because in that case the left and right
currents have same moding and our automorphism is independent from moding. When
the left and right sectors have different twists, boundary state itself carries
``twist''. The boundary condition for this case can be written similarly as the
same-twist case, if we expressed it in terms of currents themselves, not of the
modes of them. However, the boundary state has to be constructed like a
spin-field for the string on orbifold.

The final comment goes to the case of non-compact groups. The theory of
symmetric space tells us that each pair $(G,H)$ accompanies a non-compact
sibling $(G^*,H)$ in such a way that $H$ is a maximal compact subgroup of a
non-compact simple group $G^*$.\footnote{This is also called ``duality''
\cite{KN}.}  Our argument can be straightforwardly extended to the non-compact
group $G^*$ of this type. D-brane world-volume remains compact subgroup
manifold for the pair $(G^*,H)$, while T-duality transforms it to a non-compact
one in general.

\section*{Acknowledgement}

We would like to thank our colleagues in the Komaba particle theory group at
the University of Tokyo for stimulating discussions.


\section*{Appendix A}

Here we collect some technical details concerning the quantity
$T_G(\{\epsilon_i\})$.

Let us consider the simplest case of $A_n$. We number each simple root in the
Dynkin diagram of $A_n$ orderly, say from left to right. The important
properties of $T_{A_n}(\{\epsilon_i\})$ are
\beqa
T_{A_n}(\{1,\cdots,1\}) &=& \frac{1}{2}n(n+1)\,,\\
T_{A_n}(\{(-1)^{\delta_{i,k}}\epsilon_i\}) &=&
      - T_{A_n}(\{\epsilon_i\})
      + 2 T_{A_{k-1}}(\{\epsilon_{i=1,\cdots,k-1}\})
      + 2 T_{A_{n-k}}(\{\epsilon_{i=k+1,\cdots,n}\})\,.
\eeqa
These are enough to determine the value $T_{A_n}(\{\epsilon_i\})$.  The
following corollaries are also useful:
\beqa
T_{A_n}(\{\epsilon_i\}) &=& T_{A_n}(\{\epsilon_{n+1-i}\})\,,\\
T_{A_n}(\{(-1)^{\delta_{i,k}}\}) &=&
      T_{A_n}(\{(-1)^{\delta_{i,n+1-k}}\}) \; =\;
      \frac{1}{2}n(n+1)-2k(n+1-k)\,,\\
T_{A_n}(\{(-1)^{\delta_{i,k}+\delta_{i,l}}\}) &=&
      T_{A_n}(\{(-1)^{\delta_{i, \min(k , n+1-k) + \min(l , n+1-l)}}\})\,.
\eeqa

For the other groups, each $T_G$ can be given in terms of $T_{A_n}$'s:
\beqa
T_{B_n}(\{\epsilon_i\}) &=&
       T_{A_n}(\{\epsilon_i\}) +
       T_{A_{n-1}}(\{\epsilon_{i=1,\cdots,n-1}\}) \,,\\
T_{C_n}(\{\epsilon_i\}) &=&
       T_{A_n}(\{\epsilon_i\}) +
       \epsilon_n \left(T_{A_{n-2}}(\{\epsilon_{i=1,\cdots,n-2}\})
       +n-1\right)\,,\\
T_{D_n}(\{\epsilon_i\}) &=&
       (1+\epsilon_{n-1}\epsilon_n)
       T_{A_{n-1}}(\{\epsilon_{i=1,\cdots,n-1}\})\,,\\
T_{G_2}(\{\epsilon_i\}) &=&
       2 T_{A_2}(\{\epsilon_i\})\,,\\
T_{F_4}(\{\epsilon_i\}) &=&
       \left(T_{A_2}(\{\epsilon_{i=1,2}\})+1\right)
       \left(T_{A_2}(\{\epsilon_{i=3,4}\})+4\right) - 4 \,,\\
T_{E_6}(\{\epsilon_i\}) &=&
       (1+\epsilon_6)T_{A_5}(\{\epsilon_{i=1,\cdots,5}\}) \nonumber\\ && +
       \epsilon_6 \left(T_{A_2}(\{\epsilon_{i=1,2}\})-1\right)
       \left(T_{A_2}(\{\epsilon_{i=4,5}\})-1\right) +
       \epsilon_1\epsilon_3\epsilon_5(1+\epsilon_6) \,,\\
T_{E_7}(\{\epsilon_i\}) &=&
       (1+\epsilon_7) \left[ T_{A_6}(\{\epsilon_{i=1,\cdots,6}\}) +
       \epsilon_1\epsilon_3 \left(T_{A_2}(\{\epsilon_{i=5,6}\}) +
       \epsilon_4\epsilon_6\right)\right] \nonumber\\ && +
       \epsilon_4\epsilon_6 T_{A_2}(\{\epsilon_{i=1,2}\}) +
       \epsilon_7 \left(T_{A_2}(\{\epsilon_{i=1,2}\})-1\right)
       \left(T_{A_3}(\{\epsilon_{i=4,\cdots,6}\})-1\right)\,,\\
T_{E_8}(\{\epsilon_i\}) &=&
       (1+\epsilon_8) T_{A_7}(\{\epsilon_{i=1,\cdots,7}\}) +
       \epsilon_8 \left(T_{A_2}(\{\epsilon_{i=1,2}\})-1\right)
       \left(T_{A_4}(\{\epsilon_{i=4,\cdots,7}\})-1\right) \nonumber\\ && +
       \epsilon_1\epsilon_3\epsilon_5(1+\epsilon_8) +
       \left[ \epsilon_1\epsilon_3\epsilon_6(1+\epsilon_8)
       (1+\epsilon_4+\epsilon_5) +
       \epsilon_4\epsilon_6 T_{A_2}(\{\epsilon_{i=1,2}\}) \right]
       (1+\epsilon_7) \nonumber\\ && +
       \epsilon_7\left[
       (\epsilon_3 + \epsilon_3\epsilon_4 + \epsilon_3\epsilon_4\epsilon_5)
       \epsilon_1(1+\epsilon_8) +
       T_{A_2}(\{\epsilon_{i=1,2}\})T_{A_2}(\{\epsilon_{i=4,5}\})
       \right. \nonumber \\ && \left. \; +
       \epsilon_2\epsilon_3\epsilon_5(1+\epsilon_1)(1+\epsilon_8) +
       \epsilon_3\epsilon_5(1+\epsilon_8) +
       T_{A_2}(\{\epsilon_{i=4,5}\})\epsilon_8 +
       \epsilon_4\epsilon_6\epsilon_8 \right] \nonumber\\ && +
       \epsilon_4\epsilon_6\epsilon_8 \,,
\eeqa
where we have used the same numbering convention of simple roots as
ref.~\cite{S}.

In general, different sets of $\{\epsilon_i\}$ may give a same value of $T_G$.
In this case corresponding subgroups are same as a group, although the
directions are different. To get all independent values ({\it i.e.} all
different $H$) of $T_G$ it is sufficient to examine $\epsilon_i =
(-1)^{\delta_{i,k}}$, even for which degeneracy still exists.

\section*{Appendix B}

Here we prove Lemma~1.
\begin{enumerate}
\item It immediately follows from $\tau(\tau(E^{\alpha}))=E^{\alpha}$.
\item For $\alpha,\beta,\alpha+\beta\in\Delta_+$, the relation $[E^{\alpha} ,
E^{\beta}] = N_{\alpha,\beta} E^{\alpha+\beta}$ is transformed by $\tau$ to
\beq
[\varepsilon_{\alpha} E^{\tau(\alpha)} , \varepsilon_{\beta} E^{\tau(\beta)} ]
= N_{\alpha,\beta} \varepsilon_{\alpha+\beta} E^{\tau(\alpha)+\tau(\beta)}
\eeq
which should be compared with $[E^{\tau(\alpha)} , E^{\tau(\beta)}] =
N_{\tau(\alpha),\tau(\beta)} E^{\tau(\alpha)+\tau(\beta)}$. Then we have
\beq
\varepsilon_{\alpha+\beta} = \varepsilon_{\alpha} \varepsilon_{\beta}
\frac{N_{\tau(\alpha),\tau(\beta)}}{N_{\alpha,\beta}}\,.\label{NN}
\eeq
By taking $\beta=\tau(\alpha)$ the relation $N_{\alpha,\beta} = -
N_{\beta,\alpha}$ leads us to the desired result.
\item Jacobi identity for $E^{\alpha}$, $E^{\beta}$ and $E^{\tau(\alpha)}$
leads to $N_{\alpha,\beta}N_{\alpha+\beta,\tau(\alpha)} +
N_{\beta,\tau(\alpha)}N_{\beta+\tau(\alpha),\alpha} = 0$, then
\beq
\frac{N_{\tau(\alpha),\tau(\beta)}}{N_{\alpha,\beta}}
\frac{N_{\tau(\alpha+\beta),\tau(\tau(\alpha))}}{N_{\alpha
+\beta,\tau(\alpha)}} = 1 \,.
\eeq
Using the relation (\ref{NN}) this gives the desired result.
\end{enumerate}


\newpage

\begin{thebibliography}{99}

\bibitem{P} J.~Polchinski, ``TASI Lectures on D-Branes'',
            NSF-ITP-96-145, {\tt hep-th/9611050}.
\bibitem{OOY} See for recent one: H.~Ooguri, Y.~Oz and Z.~Yin,
              Nucl.~Phys.\ {\bf B477} (1996) 407.
\bibitem{HK} S.~Hirano and Y.~Kazama, UT-Komaba/96-26, {\tt hep-th/9612064}.
\bibitem{CS} C.~Klim\v{c}\'{\i}k and P.~\v{S}evera, CERN-TH/96-254,
             {\tt hep-th/9609112}.
\bibitem{GW} D.~Gepner and E.~Witten, Nucl.~Phys.\ {\bf B278} (1986) 493.
\bibitem{I} N.~Ishibashi, Mod.~Phys.~Lett.\ {\bf A4} (1989) 251.
\bibitem{G} M.R.~Gaberdiel, Nucl.~Phys.\ {\bf B471} (1996) 217.
\bibitem{CLNY} C.G.~Callan, C.~Lovelace, C.R.~Nappi and S.A.~Yost,
               Nucl.~Phys.\ {\bf B293} (1987) 83;
	       {\it ibid.\/}\ {\bf B308} (1988) 221.
\bibitem{PC} J.~Polchinski and Y.~Cai, Nucl.~Phys.\ {\bf B296} (1988) 91.
\bibitem{S} R.~Slansky, Phys.~Rep.\ {\bf 79} (1981) 1.
\bibitem{GO} P.~Goddard and D.~Olive, Int.~J.~Mod.~Phys.\ {\bf A1} (1986) 303.
\bibitem{K} E.~Kiritsis, Nucl.~Phys.\ {\bf B405} (1993) 109.
\bibitem{AAL} E.~\'{A}lvarez, L.~\'{A}lvarez-Gaum\'{e} and Y.~Lozano,
              Phys.~Lett.\ {\bf B336} (1994) 183.
\bibitem{H} S.~Helgason, ``Differential Geometry, Lie Groups, and Symmetric
            Spaces'' (Academic Press, 1978).
\bibitem{KS} Y.~Kazama and H.~Suzuki, Nucl.~Phys.\ {\bf B321} (1989) 232;
             Phys.~Lett.\ {\bf B216} (1989) 112.
\bibitem{KN} S.~Kobayashi and K.~Nomizu, ``Foundations of Differential
             Geometry'' vol.~II (John Wiley and Sons, 1969).
\end{thebibliography}
\end{document}